\documentclass[twoside]{LCWS11}
\usepackage[latin1]{inputenc}
\usepackage[dvips]{graphicx,epsfig,color}
\usepackage{wrapfig,rotating}
\usepackage{amssymb,amsmath,array}
\usepackage{cite}
\begin{document}
\title{
Depolarization in the ILC Linac-to-Ring Positron
Beamline} 
\author{Valentyn Kovalenko$^1$, Gudrid Moortgat-Pick$^{1,2}$, Sabine Riemann$^3$, Andriy Ushakov$^1$.
\thanks{Talk was presented at International Workshop on Future Linear Colliders (LCWS) 2011 in Granada, Spain, 26-30 September.}%
\vspace{.3cm}\\
1- II Institute for Theoretical Physics, University of Hamburg \\
Luruper Chaussee 149, Hamburg, D-22761 - Germany
\vspace{.1cm}\\
2- Deutsches Elektronen-Synchrotron - DESY \\
Notkestrasse 85, Hamburg, D-22607 - Germany
\vspace{.1cm}\\
3- Deutsches Elektronen-Synchrotron - DESY \\
Platanenallee 6, Zeuthen, D-15738 - Germany\\
}

\maketitle
\vspace{-65mm}
\hfill
DESY-12-017
\vspace{65mm}

\begin{abstract}
To achieve the physics goals of future Linear Colliders, it is important that electron and positron beams are polarized. The positron source planned for the International Linear Collider (ILC) is based on a helical undulator system and can deliver a polarised beam with $|Pe^+|\geq60$\%. To ensure that no significant polarization is lost during the transport of the electron and positron beams from the source to the interaction region, spin tracking has to be included in all transport elements which can contribute to a loss of polarization. These are the positron source, the damping ring, the spin rotators, the main linac and the beam delivery system. In particular, the dynamics of the polarized positron beam is required to be investigated. The results of positron spin tracking and depolarization study at the Positron-Linac-To-Ring (PLTR) beamline are presented.
\end{abstract}

\section{Introduction}

\begin{figure}[h!]
\centerline{\includegraphics[width=1\linewidth]{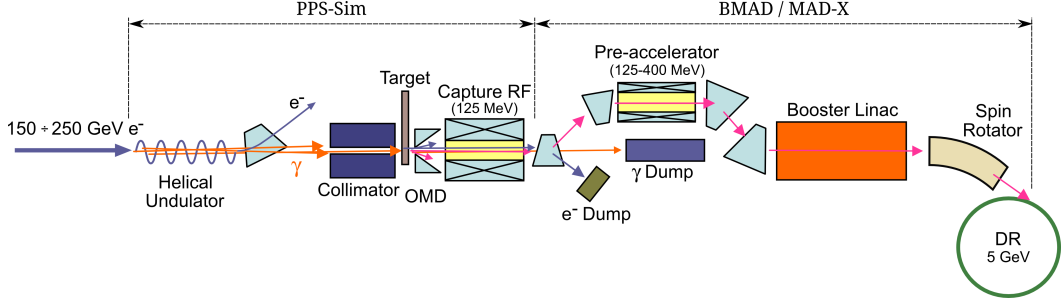}}
\caption{Positron Source Layout (RDR)}\label{Fig:PSlayout}
\end{figure}

The full physics potential of the ILC can be maximized only by using polarized electron and positron beams~\cite{Gudrid}. One scheme for the production of polarized positrons was proposed by Michailichenko and Balakin in 1979~\cite{Michailichenko}. The method is based on a two-stage process, where at the first stage the circularly polarized photons are generated in a helical electromagnetic field and then, at the second stage, these photons are converted into positrons and electrons in a thin target. The circular polarization of the photons is transferred to longitudinal polarization of the electrons and positrons. The undulator scheme of positron production has been chosen as the baseline for ILC (see Figure \ref{Fig:PSlayout})(RDR baseline). In order to properly orient and preserve the polarization of both beams up to the IP, the beam polarization must be manipulated. The spin orientation of the positrons (and electrons) has to be rotated from the longitudinal into the vertical direction before the DR and vice versa after the DR by means of spin rotators formed by  series of dipoles and solenoids. To ensure that no significant polarization is lost during the transport of the beams, precise spin tracking has to be included in all transport elements which could contribute to a loss of polarization.

\section{Spin Rotator Parameters}

\begin{figure}[h!]
\centerline{\includegraphics[width=0.25\linewidth]{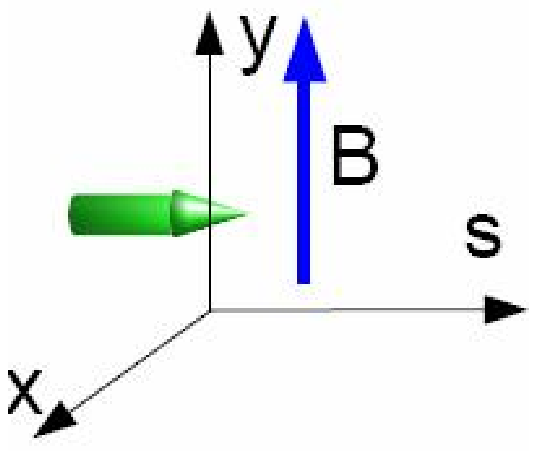} a) $~~~~~~~$ \includegraphics[width=0.25\linewidth]{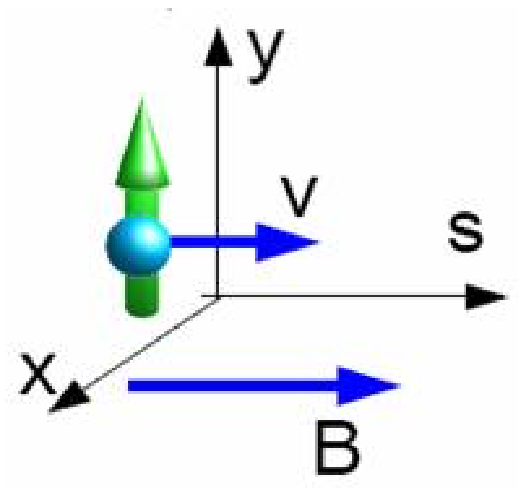} b)}
\caption{Spin (green arrows) precession in the magnetic field of a dipole (a) and a solenoid (b).}
\label{Fig:dipsol}
\end{figure}

The longitudinal polarization of the positrons is generated at the source and has to be preserved prior to the DR. To preserve polarization in the DR it is required to change the direction of the spin from longitudinal into vertical, i.e. to be parallel or anti-parallel to the magnetic field. Building blocks of the Positron-Linac-To-Ring (PLTR) beamline are bending magnets and a solenoid. In a dipole field normal to the direction of particle motion, the spin precesses around the direction of the magnetic field as shown in Figure \ref{Fig:dipsol}a. This is used to rotate the spin from the longitudinal direction to the transverse horizontal direction. This rotation of the spin by 90 degrees requires a field integral of 2.3\,Tm at 5\,GeV. The rotation to the vertical direction is done using a solenoid. The spin precesses around the axis of the solenoid field which is parallel to the motion of the particles (see Figure \ref{Fig:dipsol}b). At beam energies of 5\,GeV  the spin rotation from the horizontal to the vertical plane requires a solenoid field integral of 26.18\,Tm (see eq. $\eqref{eq:3}$).

The spin precession angle, $\varphi$, in a bending arc (dipoles) is given by
\begin{equation}
\varphi = G\gamma\theta
\label{eq:1}
\end{equation}
where $\theta$ is the bending arc angle, $G$ = 0.001159652 is the anomalous magnetic moment of the electron. $\gamma=E/m_{0}c$, where $E$ is the particle energy, $m_{0}$ is the electron rest mass and $c$ is speed of light. To rotate the spin vectors in the horizontal plane by $n \cdotp 90^{\circ}$ from the longitudinal direction, a total bending angle of $\theta = n\cdotp7.929^{\circ}$ is required (\textit{n} is an odd integer) for a 5\,GeV beam.
 
The spin rotation angle caused by a solenoid is given:
\begin{equation}
\theta_{spin}\approx \frac{B_z\cdotp L_{sole}}{B\rho}=2\theta_{orbit}
\label{eq:2}
\end{equation}
where $B_z$ is the longitudinal solenoid field, $L_s$ is the length of solenoid, and $B\rho$ is the magnetic rigidity. To rotate the spin vector of 5 GeV polarized positrons through an angle $\varphi=90^{\circ}$ a solenoid magnetic field is required:
\begin{eqnarray}
\int B_z dz=\frac{p\varphi}{e(1+G)}=26.18\;\mathrm{T\;m}
\label{eq:3}
\end{eqnarray}

\begin{figure}[h!]
\begin{minipage}[h]{0.45\linewidth}
\center{\includegraphics[width=1\linewidth]{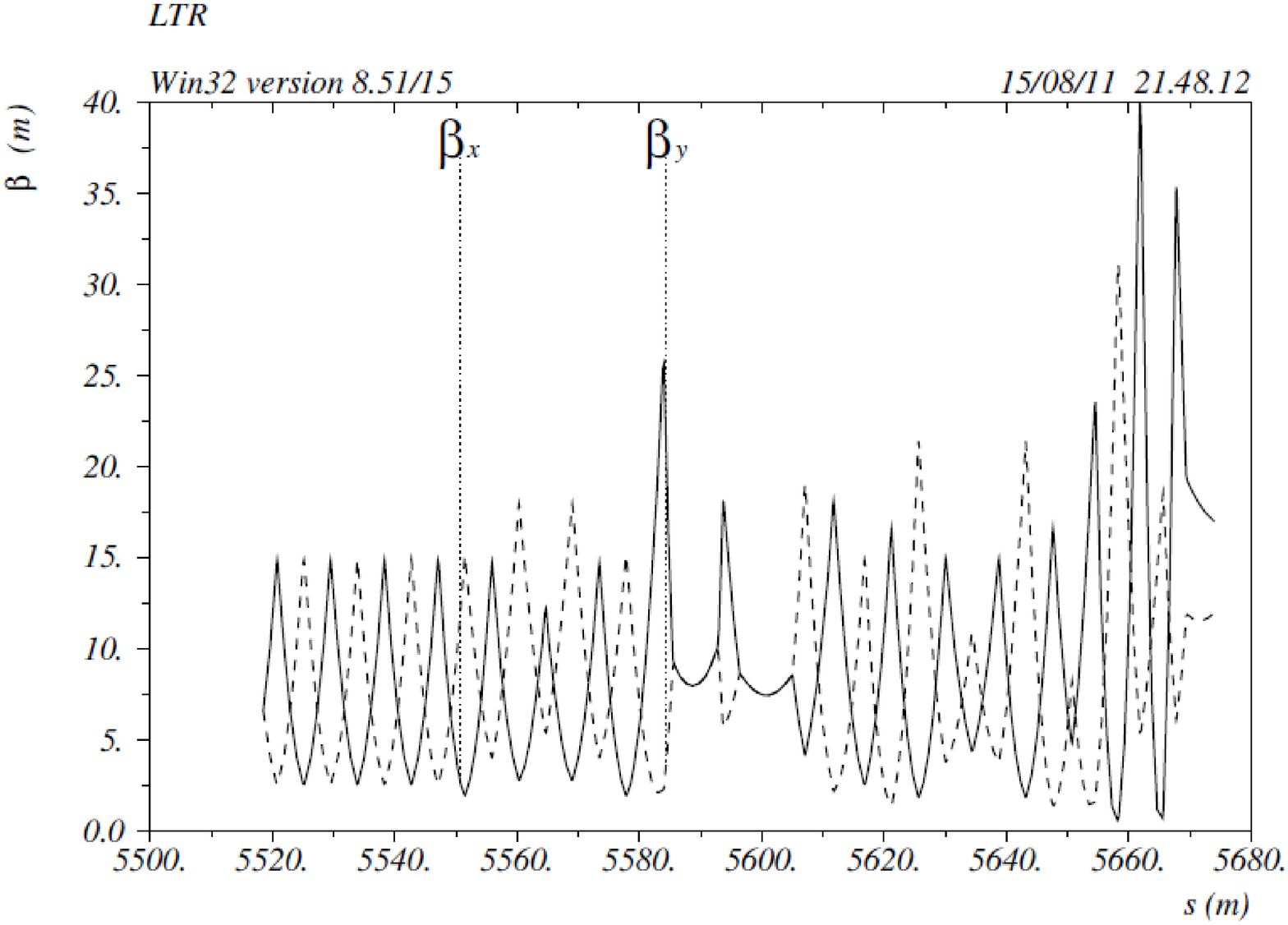} \\ a)}
\end{minipage}
\hfill
\begin{minipage}[h]{0.45\linewidth}
\center{\includegraphics[width=1\linewidth]{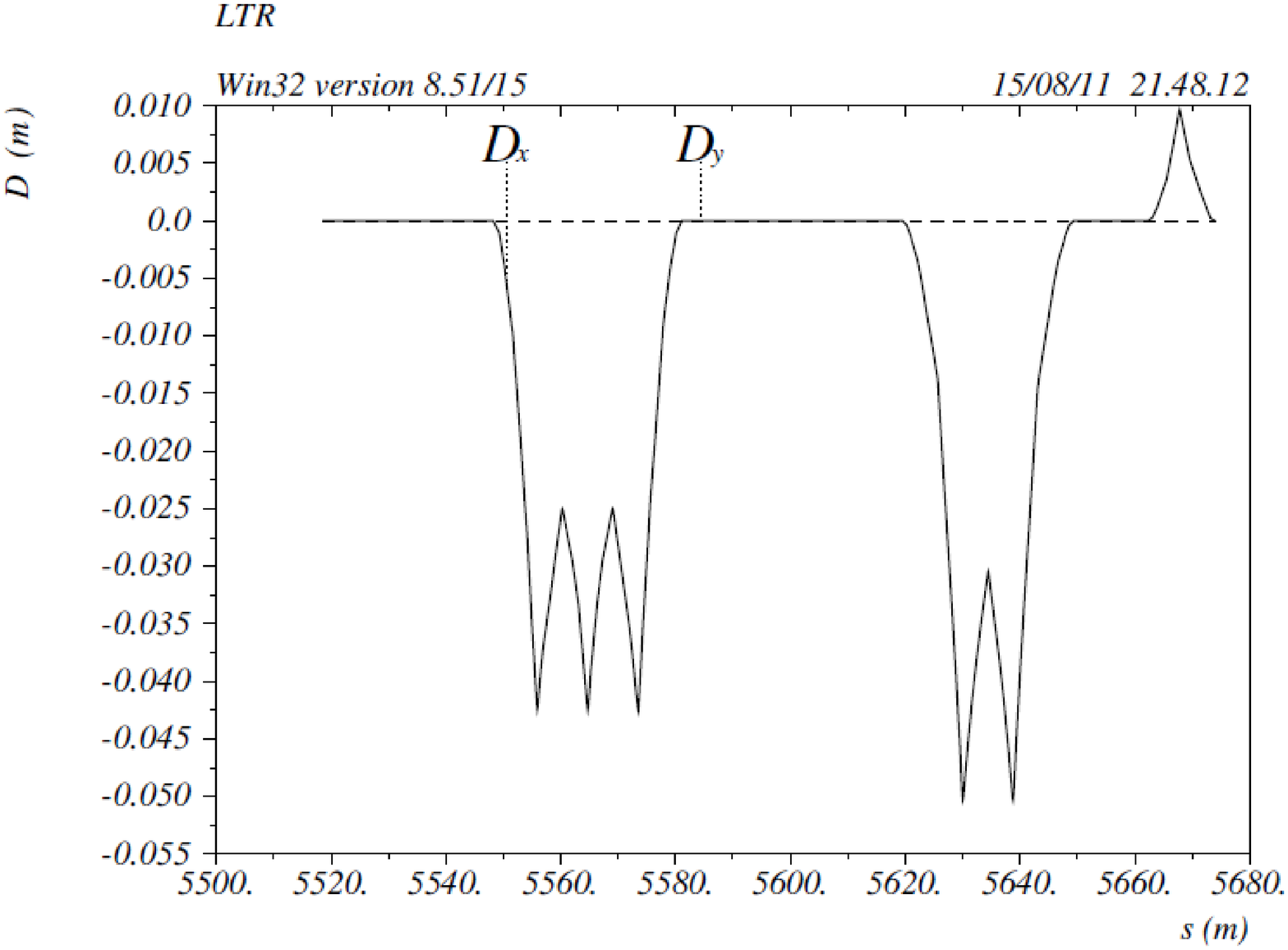} \\ b)}
\end{minipage}
\caption{$\beta$ functions (a), and dispersion (b) along the PLTR beamline.} 
\label{Fig:3}
\end{figure}

\begin{figure}[h!]
\center{\includegraphics[width=1\linewidth]{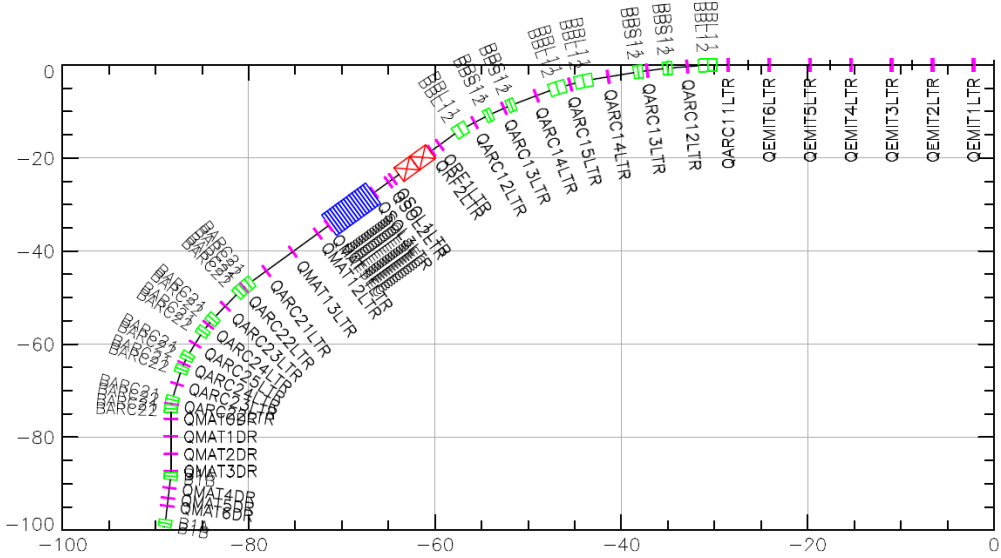}}
\caption{Floor plan of the PLTR section.} 
\label{Fig:4}
\end{figure}

If the momentum compaction factor $R_{56}$ (the ratio of the relative change in path length to the relative difference in momentum) is larger than zero the bending magnets generate bunch decompression. It leads to lengthening of the bunch. Hence, energy compression in the PLTR line is necessary to meet the DR acceptance that is restricted to an energy spread of $\Delta \gamma/\gamma=1\%$, a bunch length of 34.6\,mm and a normalized beam emittance of $\varepsilon_x+\varepsilon_y\leq 0.09$\,m-rad. The energy compression can be realized by manipulating the linac RF phase. The energy change within the bunch is given by $d\gamma/dz=eE \sin\varphi_{RF}/(mc^2)$~\cite{Batygin} where $E$ is the amplitude of the RF field, and $\varphi_{RF}$ is the RF phase of the particle. The RF voltage to transform the beam is
\begin{equation}
\ EL_{RF}=\frac{mc^2}{e}\frac{\Delta\gamma}{2\sin(\pi l_b/\lambda)}
\label{eq:4}
\end{equation}
where $L_{RF}$ is the length of the RF structure and $l_{b}$ is the bunch length at the exit of the arc and defined as $l_{b}=l_{b,0}+R_{56}\frac{\Delta\gamma}{\gamma}$  ($l_{b,0}$ is bunch length in front of the PLTR line). The first PLTR arc consists of four FODO cells and 8 bending magnets placed between quadrupoles. A total bending angle $\theta = n\cdotp7.929^{\circ}=55.5^{\circ}$ is chosen in the design. The solenoid and the RF structure are located after the arc. Altogether, the elements of the PLTR beamline perform spin rotation and energy compression. The final part of the PLTR system turns the beam by $34.5^{\circ}$ and matches the Courant-Snyder parameters at the DR injection. It should be noted that the PLTR lattice used here to study the depolarization was designed by Nosochkov and Zhou~\cite{nos+zhou}. The structure is dispersion-free; it starts with a zero dispersion region and ends with a zero dispersion region as shown in Figure \ref{Fig:3}b. It helps to avoid emittance growth due to energy spread. The dispersion is suppressed by $\pi/\mu$ FODO cells with bending magnets if $\pi/\mu=integer$~\cite{Chao}, where $\mu$ is the phase advance per cell. Choosing $\mu=90^{\circ}$ gives $\pi/\mu=2$ and makes the structure shorter. Therefore the arc contains $360^{\circ}/90^{\circ}=4$ FODO cells (see floor plan in Figure \ref{Fig:4}).

\section{Spin Tracking Simulation and Results}

The depolarization study of the PLTR beamline was done using the BMAD~\cite{BMAD} code. Spin tracking has been implemented in BMAD by Jeff Smith~\cite{Smith} using the spinor-quaternion transfer map method. The spin motion is described by the T-BMT equation. In 2-component spinor notation it can be written as
\begin{equation}
\frac{d}{dt}\Psi=-\frac{i}{2}(\mathbf{\sigma}\cdot\mathbf{\Omega})\Psi
\label{eq:5}
\end{equation}
where $\mathbf{\sigma}$ are Pauli matrices and $\mathbf{\Omega}$ is the spin precession vector. The solution can be written as
\begin{equation}
\Psi=(a_0\mathbf{1}_2-i\mathbf{a}\cdot\mathbf{\sigma})\Psi_i=\mathbf{A}\Psi_i
\label{eq:6}
\end{equation}
with the spinor $\Psi=(\psi_1,\psi_2)^T$ ($\psi_1$ and $\psi_2$ are complex numbers). The matrix $\textbf{A}$ is called quaternion and describes the transfer map for each element. Tracking through any element is achieved via the application of these quaternions in sequence. This results in very fast tracking times.

\begin{figure}[h!]
\center{\includegraphics[width=1\linewidth]{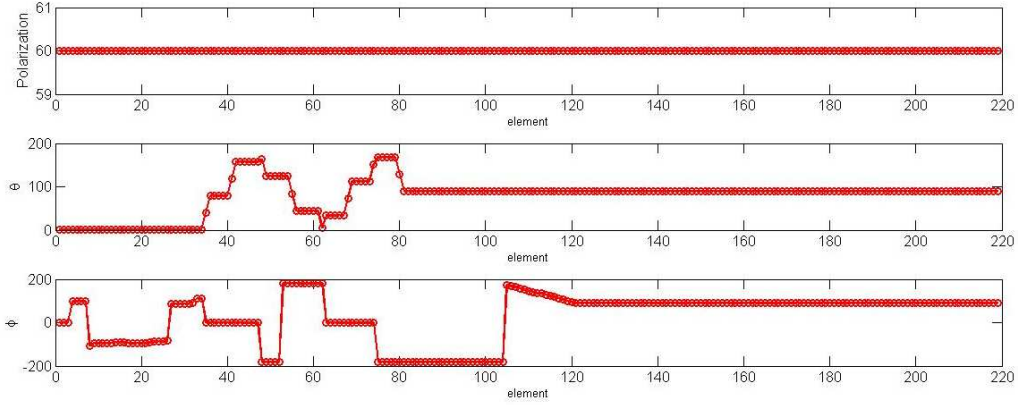}}
\caption{Polarization and average spin direction along the beamline.
Ideal Beam: emittance, bunch size, energy spread are very small (10 orders less than expected).}
\label{Fig:5}
\end{figure}

Let us consider now the idealized case when emittance, bunch size and energy spread of the beam are very small, around 10 orders less than expected to be in reality. The results of spin tracking of this ``ideal'' beam through the PLTR line are presented in Figure \ref{Fig:5}.
 
The beam energy spread yields polarization loss, $\delta(\frac{\Delta P}{P})=1-\cos(\delta\varphi)$~\cite{Depol_in_SLC}, where $\delta\varphi=\varphi(\delta\gamma/\gamma)$ is the deviation of the spin rotation angle from its nominal value $\varphi$ because of energy deviation $\delta\gamma$. Hence, a large depolarization is observed for larger beam energy spread. According to \cite{Montague} the relative depolarization is $1-\dfrac{\langle P_z \rangle}{P_0}$. The mean polarization, $\langle P_z \rangle$ is given by 
\begin{equation}
\langle P_z \rangle =P_0\exp{(-\frac{1}{2}(G\gamma\alpha\sigma_E)^2)}
\label{eq:7}
\end{equation}
where $G$ is the anomalous momentum, $\alpha$ is the arc bending angle, and $\sigma_E$ is the rms energy spread. Taking into account equation~$\eqref{eq:7}$ we expect for fixed bending angle and small energy spread a small depolarization as demonstrated by numerical simulations shown in Figure \ref{Fig:5}. It can also be seen from Figure \ref{Fig:5} (polar angle $\theta$ and azimuthal angle $\phi$ of the spin) that the arc with bending magnets provides spin rotation in the median plane and the 
solenoid performs spin rotation into vertical direction so the beam has transverse polarization at the end of spin rotator ($\theta$ and $\phi$ are equal $90^{\circ}$).

\begin{figure}[h!]
\center{\includegraphics[width=1\linewidth]{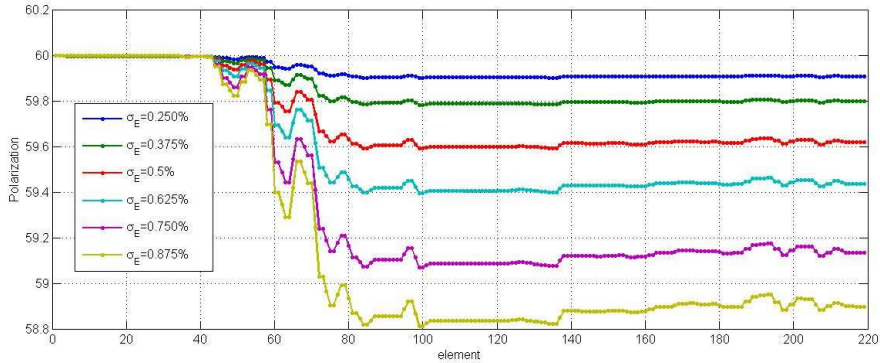}}
\caption{Impact of bunch energy spread on beam polarization.}
\label{Fig:6}
\end{figure}

We assume that the energy distribution within the bunch is Gaussian 
\begin{equation}
\frac{dN}{d(\delta\gamma)}=\frac{1}{\sqrt{2\pi}\sigma_E}\exp[-\frac{(\delta\gamma)^2}{2\sigma^2_E}]
\label{eq:8} 
\end{equation}
with the rms value $\sigma_E$.\\
\begin{wrapfigure}{r}{0.5\columnwidth}
\centerline{\includegraphics[width=0.45\columnwidth]{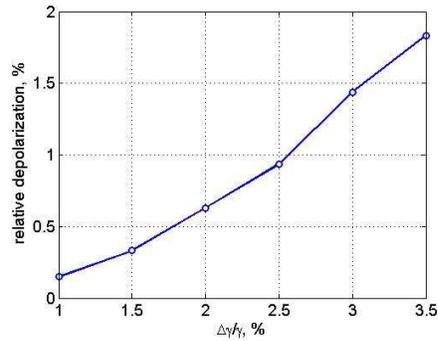}}
\caption{Relative depolarization due to energy spread}\label{Fig:7}
\end{wrapfigure}
The full width energy spread of the transformed bunch is $\Delta\gamma/\gamma$ and the corresponding rms energy spread is approximately $\sigma_E/\gamma\approx(1/4)\Delta\gamma/\gamma$. The impact of the bunch energy spread on beam polarization is shown in Figure \ref{Fig:6}. Here we can observe that an increase of the bunch energy spread leads to depolarization of the beam. Equation $\eqref{eq:7}$ tells us that polarization depends exponentially on the energy spread. The beam depolarization due to energy spread is shown in Figure \ref{Fig:7}. For $\Delta\gamma/\gamma=3.5\%$ the depolarization reaches $1.8\%$. A small bending angle of the spin rotator is better to reduce the depolarization effect for a fixed energy spread.

\section{Conclusions}

A positron spin tracking and depolarization study has been performed for the ILC Positron Linac To Ring beamline. The PLTR arc consisting of four FODO cells and bending magnets placed between quadrupoles provides rotation of the spin in the horizontal plane. Then the solenoid which is a part of PLTR line rotates the spin into vertical direction. Numerical simulations carried out using the BMAD code show a relative depolarization of 1.8\% for 3.5\% initial positron energy spread and $55.5^{\circ}$ bending angle. To reduce the depolarization effect the bending angle of the PLTR line should be smaller.

\section{Acknowledgments}
We would like to thank the organizers of the LCWS'11 for this fruitful and encouraging workshop and for hospitality.

This work is supported by the German Federal Ministry of Education and Research, Joint Research Project R\&D Accelerator ``Spin Management'', contract number 05H10GUE.

\section{Bibliography}


\begin{footnotesize}


\end{footnotesize}


\end{document}